# Calibration of photomultiplier using UVLED

Sungrae Im[1*], Bokkyun Shin[2], Ikeda Daisuke[3], Stan Thomas[4] and ByungGu Cheon[1]

1. Dept. of Physics, Hanyang University

2. LaCosPa, National Taiwan University

3. Earthquake Research Institute, Tokyo University

4. Dept. of Physics & Astronomy, University of Utah

## Abstract

Detector calibration is very important for the long-term operation. For the purposes of simple and precise calibration, we developed a new portable calibration source using UVLED to calibrate the fluorescence detector of the Telescope Array experiment (TA). The UVLED is light (less than 1 kg) and its setup is quick and easy. Therefore, a single source will be used instead of the calibration sources that TA currently uses.

The UVLED emits 369 nm wavelength and ~70 pJ of photons which are calibrated with 5% accuracy using photodiode. It has a particular unit, which is an internal heater, to keep the temperature constant and avoiding temperature dependency on light yield. The first result of calibration for 24 detectors is xx photons/faced count for 369 nm photons with 6% uncertainty. It is in agreement with the result of the current calibration sources of TA.

# I. INTRODUCTION

The Telescope Array (TA), located in Utah, USA, is an experiment carried out to conduct research on the Ultra High Energy Cosmic Ray (UHECR). The UHECRs are most energetic particles of the universe with energies greater than $10^{18}$ eV and their existence is the mystery in modern astrophysics: their birth and the acceleration process must be related to the most energetic phenomena in the universe. To detect UHECR, the TA experiment utilizes two types of detectors for "air-shower," which is a generation of considerable number of secondary particles through an interaction between UHECR and atmospheric molecule. One type is an array by 507 of surface detectors (SDs) with 1.2 km spacing in an area of 700 km$^2$, and the other type is composed of three stations of fluorescence detectors (FDs), overlooking the sky above the SDs. These detectors are operated in "hybrid mode" to observe the air-shower. [1].

We describe a new method of on-site calibration of FD stations using the UVLED. The energy from FD is precise due to the tracking of longitudinal shower development. It requires four calibration parameters such as PMT gain. We developed a new calibration source using the UVLED to calibrate and monitor the FDs. The UVLED is portable (less than 1 kg) and consists of stable light yield using a LED lamp. In this paper, we estimate the calibration factor for 6,144 photo multifiber tubes (PMTs) of two stations, BRM, and LR, which are located south-east and south-west of the TA site. This UVLED will cover absolute calibration total of six stations of not only TA but also TAx4 and TALE [2].

# II. EXPERIMENTS AND DISCUSSION

## II -1. Current Calibration and Monitoring of FD

TA consists of three stations of FD, and each of them is composed of 12 or 14 telescopes. The FD telescope consists of 256 the UV sensitive PMTs and an imaging mirror.

TA has been operated for 10 years in a wild environment. Therefore, there are monitoring and calibration tools for PMT response such as the absolute, relative gain of PMT [3], and long-term response [4] using various calibration source.

To calibrate FD, the TA uses four calibration items with the help of three calibration sources. The final calibration factor G with four items is described as

$$G = G_0 \times G_1 \times G_2 \times G_3.$$

The $G_0$ is absolute calibration using CRAYS factor by the ratio of the number of hits of photons to the digital counts of electronics. The CRAYS is a system of calibration used at the Institute of the Cosmic Rays Research, Japan, which consists of 337.1 nm wavelength pulse laser and 99.9% purity nitrogen-filled chamber. The PMT is set on the chamber which is at an orthogonal direction of the laser. Thus, 90-degree scattered photons illuminate the PMT photocathode through a 16 cm$^2$ window which is located at a distance of 4 cm perpendicular to the beam axis. The number of scattered photons is estimated by Monte-Carlo. The PMT response is digital count recorded by FD electronics. Using the CRAYS, we calibrated 50 PMTs of the so-called 'standard PMT' in ICRR and attached a small light pulser to the so-called YAP. Calibrated PMTs are sent to the TA-site in Utah and set as 2~3 PMTs per telescope of the Long Ridge and Black Rock Station.

The absolute calibration by using only CRAYS for all PMT is limited by time. Thus, we calibrated all PMTs by the relative comparison with 50 CRAYS and calibrated PMTs using Xenon flasher (XF). We assigned this relative as a $G_1$ item. The XF is set on the center of the mirror and shot photons uniformly to PMT camera surface. We estimated relative calibration factor relative to and in comparison with standard PMTs. To check long-term stability, we monitored standard PMT response by the YAP ($G_2$); currently, we did not observe the aging effect. [3]. The last item $G_3$ is the measurement of temperature dependency. This is achieved using the YAP in an incubator, tuning temperature -20 to 40 °C. The result of temperature coefficient is -0.727 %/°C.

## II -2. Calibration using UVLED

We developed a new light source named UVLED to calibrate FD more precisely and simply. As described in section 2, TA has four items of calibration. The stable and absolutely calibrated UVLED can cover item $G_0$-$G_2$ using a single source. So, 11% uncertainty from $G_0$ to $G_2$ is merged as one item of uncertainty by UVLED.

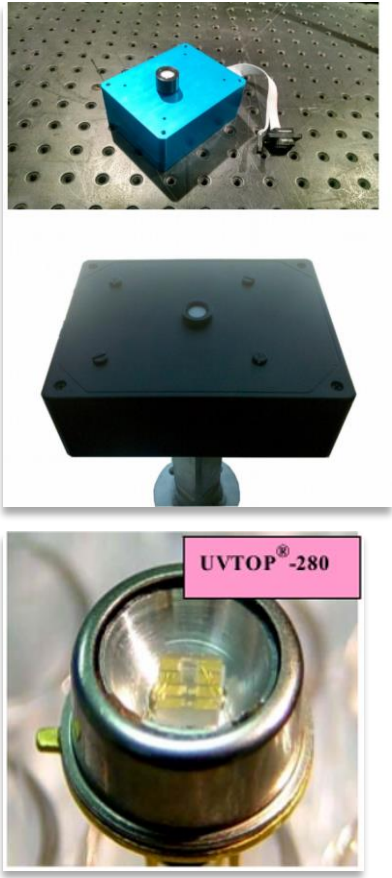

**Fig. 1-(a), (b) Image of UVLED**

## II -2-1. The UVLED

The UVLED, shown in Fig. 1, contains UVLED lamp (UVTOP355, SETi), Teflon defuser, and control electronics. The UVLED emits 369 +/- 5 nm wavelength of a light pulse at a repetition rate of 10 Hz. The energy of pulse, in a condition of 1us pulse width and 45 ℃, is 70.47 pJ. This is measured using a laser photodiode (RjP-465, LaserProbe). The photons are uniformly distributed by the Teflon defuser.

The light intensity of light source depends on temperature. However, the temperature range of the TA-site, a desert in Utah, ranges between -20 ℃ to 40 ℃. To avoid this temperature dependency, the UVLED has an inner heater which keeps the temperature of the UVLED circuit constant at 45 ℃, so the UVLED can emit the same number of photons temperature independently.

For FD calibration run, The UVLED is connected to the mirror in the same distance (around 2 m) from the PMT camera surface. The mirrors are covered by a black cloth to avoid reflection between mirror and PMT camera surface. The photons from UVLED illuminate all the PMTs uniformly.

We performed 1,000 events of UVLED pulse shots data of each telescope. Standard telescope BR 6 and LR 4 are taken twice for the systematic checks.

## II -2-2. UVLED run

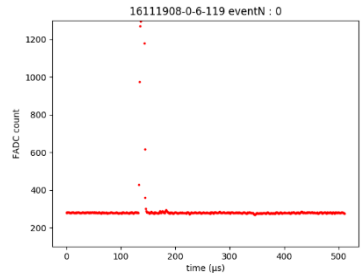

**Fig. 2 time window of UVLED - Each of the events consists of 512 bins with 0.1 μs width. The x-axis is a time in us and y-axis is digital count (FADC count) equivalent of charge from PMT recorded by FD electronics.**

In Fig. 2, we plot waveform of a UVLED event in a single PMT. The charge of the UVLED signal ($S_{event}$) is the calculated integration of FADC count subtracted by the pedestal ($S_{ped}$),

$$S_{event} = \sum_{peak-1\mu s}^{peak+2\mu s} \left(FADC_i - S_{ped}\right).$$

Peak is the bin having maximum FADC count in one event. $S_{ped}$ is the pedestal level, which is the average of first 100 bins.

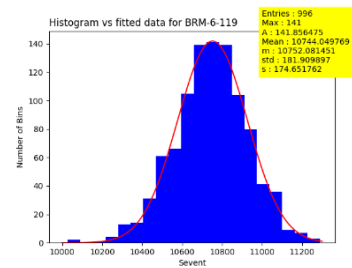

**Fig 4. Distribution of $S_{event}$ of one camera in a run. The red line is a result of a Gaussian function. The shot to shot fluctuation is about 2%.**

## II -2-3 Calibration Factor (Np/FADC)

The definition of calibration factor for 369 nm wavelength by UVLED ($G_{UVLED}$) is the number of hit photon ($NP_{PMT}$) on PMT photocathode per unit digital count ($S_{PMT}$) [FADC],

$$G_{UVLED} = NP_{PMT} / S_{PMT}.$$

The $S_{PMT}$ is calculated by the average of $S_{event}$ in a run. Fig. 4 shows a distribution of $S_{event}$ in BRM camera 6 PMT ID 0x77 which is placed at the center of PMT camera. We rejected events as cosmic muon hit, too narrow time, and unstable pedestal. The averaged charge in a PMT ($S_{PMT}$) is defined by mean of $S_{event}$ in one run. The shot to shot fluctuation is 1.8% and a total number of events is about 1000. So, this corresponds to the statistic error of $S_{PMT}$ is 0.1%.

The $NP_{PMT}$ on the PMT is estimated as

$$NP_{PMT} = NP_{LED} cos^N \theta \Omega / \pi,$$

where $NP_{LED}$ is a total number of emitted photons from UVELD measured by photodiode [5], and $\Omega$ is a solid angle between UVLED to PMT surface. The $cos^N \theta$ corresponds to angular intensity dependency, including a light source and PMT. We assigned N factor four causes of three effects: Lambert law (N = 1), light inverse-square law (N = 2), and light's orthographic projection law (N = 1).

## II -3 UVLED calibration

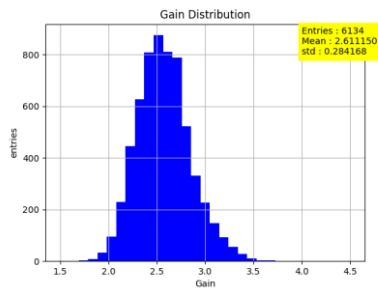

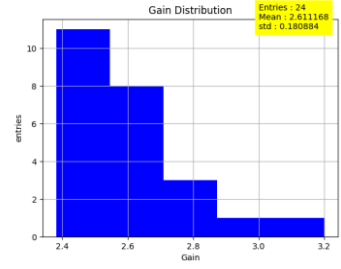

**Fig. 5-(a) Calibration factor for every PMT in BRM & LR. Average of gain is 2.61, and statistic error is 0.18%.**

**Fig. 5-(b) Calibration factor for every camera in BRM & LR. Average of gain is 2.61.**

We evaluated PMT PMT responses in a camera, distributed about +/- 10% from the average, and confirmed by the current $G_1$ factor. The average of all PMTs is 2.61 +/- 0.28 [Np/FADC]. This is consistent with the calibration factor by CRAYS, 2.82 +/- 0.31 [6]. The uncertainty is estimated not only 8% of CRAYS, but also ~6% aging effect for the 8-year uncertainty of YAP analysis. These analysis results are in agreement with the previous method.

## II-4 Systematic Uncertainty estimation

To confirm UVLED characteristics such as $NP_{LED}$ and N factors, we compared with XF calibration methods.

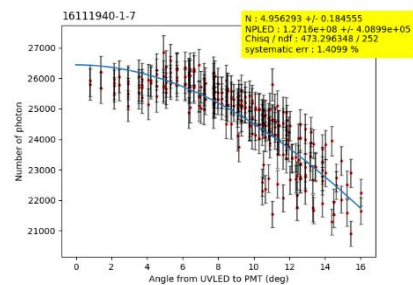

**Fig. 6 Chi-square / ndf fitting of BRM, camera 07. The horizontal axis corresponds to the angle from UVLED to PMT. The vertical axis corresponds to a number of photons received by PMT. The blue line shows fitted line with two parameters.**

Fig. 6 is the result of the UVLED event in a camera between angular position and Np of each PMT. The definition of angular photons is the angle between

UVLED; 0° and 16° correspond the center and edge of the camera. The number of photons are estimated using current G with wavelength correction, as described in section **I**. The 2% error bars are the geometrical uncertainty of the $G_1$ factor.

Fig. 6 shows the largest systematic error among 24 telescopes. The systematic error is calculated by comparison of two fitting values where N is free and N is fixed into 4.

**II -2-5 systematic uncertainty for running by run $S_{PMT}$ fluctuation.**

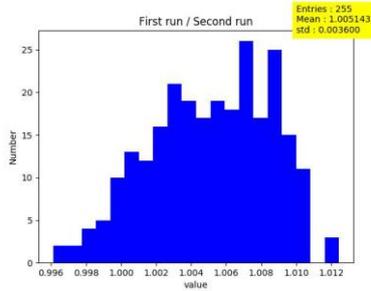

**Fig. 7 comparison of $S_{charge}$ for two runs in same PMT in LR station 04**

To confirm stability between measurements, we compared data of the first and second runs using the telescope 04 in LR. Fig. 7 shows the comparison of PMT output between first and last UVLED run of telescope 04 of LR station. The PMT output FADC are corrected by temperature dependence discussed in section **II-1**. The second run result is about 0.5% shifted from the first during the four hours of measurement.

The systematic uncertainty of UVLED calibration factor is 5.2% and each uncertainty factor is described in Table 1.

**Table 1. The systematic uncertainty of the calibration**

| Systematic uncertainty | uncertainty | Uncertainty factor |
|---|---|---|
| Absolute Np of UVLED | 5 % | The uncertainty of Photodiode [RjP-435] |
| Time passing | 0.5 % | SLED Discrepancy between first and last runs |
| Cos N factor | 1.4 % | N factor model with the fitting result |
| Total | 5.2 % | Quadruple summation above times |

### III. CONCLUSION

We study FD calibration analysis of two stations using UVLED in a TA experiment. The calibration factor evaluated from UVLED is 2.61 +/- 0.28. This result also confirms the result of current calibration methods by CRAYS, XF, and YAP within its uncertainty. We have plans to determine calibration items, confirm long-term stability, and determine individual PMT temperature coefficient ($G_2$ and $G_3$ in section **II -1**). Also, we will not only use it for TA FD telescopes, but also for additional FD telescopes, such as the TALE for lower energy CR research and TAx4 for next generation of TA.